\documentstyle[prl,aps,twocolumn]{revtex}

\begin{document}

\title{Inflation and Generalized O'Raifeartaigh SUSY models}

\author{
Arjun Berera$^{1}$
\thanks{E-mail: ab@ph.ed.ac.uk}
and Thomas W. Kephart$^{1,2}$
\thanks{E-mail: kephartt@ctrvax.vanderbilt.edu}
}

\address{
{\it $^{1}$School of Physics,
University of Edinburgh, Edinburgh EH9 3JZ, Great Britain} \\
{\it  $^{2}$Department of Physics and Astronomy, Vanderbilt University, Nashville TN 37235, USA}}

\maketitle

\begin{abstract}

Thermal inflation usually requires an inflationary potential with 
nonrenormalizable operators (NROs). We demonstrate how O'Raifeartaigh models
with or without NROs can provide thermal inflation and
a solution to the moduli problem, as well as provide SUSY breaking.
We then discuss a scenario where generalized O'Raifeartaigh potentials
(with NROs) are included in a SUGRA where the supergravity and
O'Raifeartaigh potentials provide negative and a positive contributions 
to the cosmological constant respectively. Tuning these contributions to nearly
cancel can provide the present value of the dark energy.

\end{abstract}

\medskip 

In Press Physics Letters B 2003

\medskip
\medskip


There is considerable belief that the fundamental model of
particle physics respects local and/or global supersymmetry at high energy.  
Inflationary cosmology appears to provide further support
to this expectation. Due to the ability of supersymmetry to protect
against radiative corrections, such models provide
powerful means to realize ultra-flat potentials, which 
are necessary from inflation density perturbation constraints.
However, alongside this benefit,
cosmological implementations of
supergravity and  SUSY models generally lead to
undesired particles, such as the spin 3/2 gravitino
in supergravity models \cite{gravitino} 
and various spin zero particles of mass
$\sim 10^{2-3} {\rm GeV}$ \cite{moduli}.  In particular for cosmological
inflation, whether supercooled \cite{si} or warm \cite{wi},
which end at conventional high temperature scales, 
$T \stackrel{>}{\sim} 10^{10} {\rm GeV}$,  overabundances of
unwanted SUSY particles is a real problem, sometimes termed
the moduli problem \cite{moduli,Lyth:1995hj}.  

SUSY must not survive at low energy scales, where
physics clearly is not supersymmetric, with current
limits set by particle physics experiments indicating
SUSY must break above the electroweak scale $\sim 10^3 {\rm GeV}$.
It is reasonable to expect that symmetry breaking
and more specifically SUSY breaking
has cosmological implications.  {}For example, one scenario
termed thermal inflation 
\cite{Lyth:1995hj,Lyth:1995ka,Lazarides:ja,Barreiro:1996dx,Asaka:1999xd}
uses symmetry breaking to
overcome the problem of overabundance
of unwanted particles  
created by SUSY at high temperature.
A second problem related to SUSY, for which
cosmologists are universally and anxiously 
awaiting an explanation, is the
presentday cosmological constant $\rho_{\Lambda}$.
Observation of type IA supernova data have indicated 
an accelerating universe \cite{supernova}, 
which could be explained by a cosmological
constant of $70 \%$ of the critical density, which implies
a vacuum energy component $\rho_{\Lambda} \sim 10^{-10} {\rm eV}^4$.
Recently the first year WMAP data has independently verified the
presence of a cosmological constant, finding 
$\Omega_{\Lambda} = 0.73 \pm 0.04$ \cite{wmap}.

In this Letter, we will demonstrate that generalized O'Raifeartaigh
models \cite{O'Raifeartaigh:pr}
can realize thermal inflation and solve the
presentday cosmological constant problem.
Recall that spontaneous global SUSY breaking can be accomplished by the
O'Raifeartaigh mechanism that requires at least three chiral
supermultiplets. The minimal model has a superpotential of the form
\begin{equation}
W(\phi ,\chi ,\eta )=a \chi \left[ \phi ^{2}-M^{2}\right] +m\eta \phi .
\end{equation}
SUSY is broken since the requirement 
$\frac{\partial W}{\partial \phi_{i}}=0$,
with $\phi_{i}={\phi,\chi,\eta}$,
cannot be satisfied for all three fields.  In other words
the three conditions,
\begin{eqnarray}
\phi ^{2}-M^{2}=0, 
\phi =0, 
2a\chi \phi +m\eta =0 ,
\end{eqnarray}
cannot be simultaneously satisfied.
Our purpose is to
demonstrate that within their compact structure, these models contain 
nontrivial cosmological implications.  We will begin with
a review of thermal inflation, to understand the relevant scales
necessary for such scenarios.  Generalizations of
the O'Raifeartaigh models are then presented and
solutions are derived for thermal inflation and
the presentday cosmological constant.
We then briefly discuss embedding
O'Raifeartaigh models in supergravity (SUGRA) and other
fundamental theories, as well
as particle physics implications of such models.

The thermal inflation scenario is comprised of two
phases of inflation. The first phase is the
normal one, typically motivated by GUT physics and
pictured to end, after reheating, at a high temperature
$T \stackrel{>}{\sim} 10^{10} {\rm GeV}$.  In this phase,
the large scale physics is determined, such as
density  fluctuations. The key
new feature that underlies thermal inflation is that it requires the
presence of a scalar field $\phi$, often called the
flaton, which has a symmetry breaking potential
with the properties that at high temperature, $T > V_0^{1/4}$ symmetry
is unbroken with $\phi=0$ where the scale of the potential is
$V_0^{1/4} \approx 10^9 {\rm GeV}$.  On the other hand,
at $T=0$ symmetry is broken with the minimum now at
$\phi \approx 10^9 {\rm GeV}$ and with the scalar particles
acquiring a mass $m_{\phi} \sim 10^{2-3} {\rm GeV}$.
Given such a potential, a second phase of inflation,
termed thermal inflation, commences.  In this picture,
for $T > V_0^{1/4}$ the scalar field finite temperature effective
potential locks the flaton field at $\phi = 0$ and the universe is in
a hot big bang regime.  Once $T < V_0^{1/4}$, the potential
energy of this field dominates the energy density of the
universe, thereby driving inflation, which to a good approximation is 
assumed to be an isentropic expansion.
Due to the high temperature corrections to the effective potential,
in the initial phase of thermal inflation, the scalar field
remains locked at its high temperature point, $\phi = 0$. However, 
since inflationary expansion is rapidly cooling the universe, it implies
the effective potential is evolving to its zero temperature form. 
Eventually, in what is estimated to be $\stackrel{<}{\sim} 15$ e-folds,
the scalar field VEV no longer is locked at zero, and is
able to roll down to its new minimum.   

The effect of the second phase of inflation is to lower the
temperature of the universe from $T \sim 10^9 {\rm GeV}$ to
$T \sim 10^3 {\rm GeV}$.  This alone does not solve any overabundance 
problems since the abundance ratios $n/s$ for all species remain
constant.  However, subsequent to thermal inflation the scalar field
oscillates, thereby producing scalar particles of mass
$m_{\phi} \sim 10^{2-3} {\rm GeV}$ and lighter.  These particles
eventually decay, producing a huge
increase in entropy, thereby adequately diluting the abundances of
unwanted relics.  {}Finally, in order not to affect the success of
hot big bang nucleosynthesis, the temperature after decay of
scalar particles is constrained to be above $\sim 10 {\rm MeV}$.
Note, the desired features of thermal inflation could also occur
for a continuous phase transition and a nonisentropic, warm-inflationary
type expansion, which dampens the flaton's motion during its evolution
to its new minimum \cite{wi,br}.

The details of the thermal inflation scenario outlined above 
can be found in 
\cite{Lyth:1995hj,Lyth:1995ka,Lazarides:ja,Barreiro:1996dx,Asaka:1999xd}.
The key point demonstrated in
these papers is that all the desired features of this scenario
follow, provided a potential with the properties described above is
present.  Considerable work on thermal inflation studies the
consequences of such potentials, but many fewer
works attempt to find explicit models of such potentials.
Thermal inflation is typically carried out with potentials containing
higher (\mbox{$>$}$4$) 
dimension operators that are suppressed by powers of the Planck mass.
In most studies of thermal inflation
\cite{Lyth:1995hj,Lyth:1995ka,Lazarides:ja,Barreiro:1996dx,Asaka:1999xd},
SUSY breaking is handled separately, for example,
through nonperturbative means, such as the
Affleck-Dine mechanism. Here we observe that a generalization of the
O'Raifeartaigh potential, with one term replaced by a higher dimension
operator can provide SUSY breaking, thermal inflation,
and potentially, the presentday cosmological constant. 
Aside from the compactness of this solution, another advantage
is that SUSY breaking terms are calculable at the tree level
in the renormalizable O'Raifeartaigh model, so one has
more control in model building.  {}For the generalized
O'Raifeartaigh model, loop level
calculations would diverge.  
However, the basic motivation of the higher dimension
operators is string theory which would serve to cut off all divergences
and still leaves the model with some degree of control.

To treat the cosmological moduli and cosmological constant problems,
consider the 
generalization of the O'Raifeartaigh model superpotential,
\begin{equation}
W(\phi ,\chi ,\eta )=a\chi \left[ \phi^2
-M^{2}\right] +\lambda\eta \frac{\phi ^{n+1}}{m_{Pl}^{n-1}} .
\label{genor}
\end{equation}
The $\frac{\partial W}{\partial \phi _{i}}=0$ conditions now become
\begin{eqnarray}
\frac{\partial W}{\partial \chi }=a[\phi^2
-M^{2}]=0 \nonumber \\
\frac{\partial W}{\partial \eta }=\lambda\frac{\phi ^{n+1}}{m_{Pl}^{n-1}} =0
\nonumber \\
\frac{\partial W}{\partial \phi }=2a\chi \phi +(n+1)\lambda\eta
\frac{\phi ^{n}}{m_{Pl}^{n-1}} =0 ,
\end{eqnarray}
and since these cannot be simultaneously satisfied, SUSY is broken.
To carry out the calculation of thermal inflation
and the cosmological constant, we need the Higgs
potential $V=\left( \frac{\partial W}{\partial \phi _{i}}\right)
^{*}\left(
\frac{\partial W}{\partial \phi _{i}}\right) $, which is
\begin{eqnarray}
V & = & a^{2}|\phi^2-M^{2}|^2+\lambda^2m_{Pl}^{4}
\frac{|\phi|^{2(n+1)}}{m_{Pl}^{2(n+1)}}
\nonumber \\
& + & | 2a\phi \chi +(n+1)\lambda\eta\frac{\phi^n}{m_{Pl}^{n-1}}|^2 .
\label{genorpot}
\end{eqnarray}

The first objective is to show at zero temperature this potential
has the correct features and scales for thermal inflation
and the presentday cosmological constant. 
{}For this the minimum of $V$ is required.
There is a single family parameter of minima with
$\langle \phi \rangle  \neq 0$, given by setting the 
third term in the potential to zero. This gives the
condition $\langle \eta \rangle = x \langle \chi \rangle$,
with $x= -2am_{pl}^{n-1}/(n+1)\lambda \langle \phi \rangle^{n-1}$.
A number of possibilities exist for this direction.
{}First if the flat direction is uncorrected by the
full theory, then there will be a massless
boson $b = \chi+ x \eta$.  If this boson couples
sufficiently weakly to standard model fields, it does not upset the cosmology.
In particular, although it will not thermalize, it still redshifts away.
{}For a more strongly coupled $b$, particle physics familon
limits apply \cite{Ammar:2001gi}.  If the full theory corrects
the O'Raifeartaigh potential, the mass generated
for $b$ will allow it to contribute to the dark matter density
or $b$ could generate an additional moduli problem at lower
scale.  A further implication is that corrections from
outside the O'Raifeartaigh potential could allow
the overconstrained set of conditions on the VEVs to be
relaxed in a way that
spoils the O'Raifeartaigh mechanism
and could restore SUSY.
We assume this does not
happen or if it did we would have to modify the O'Raifeartaigh
potential to again make the system overconstrained and thus
break SUSY.  

This model also will have a goldstone fermion (goldstino)
once global $N=1$ SUSY is broken.
Nevertheless, should such particles be produced, they will redshift
away like radiation.  However, the other fermions generally
will have mass and this leads to an interesting
possibility.
These fermionic components could be identified with right-handed
neutrinos, for example if the U(1) symmetry
of the generalized O'Raifeartaigh model was
identified with B $-$ L.   In this case a leptonic asymmetry
can be produced, which
can lead to baryongenesis based on the scenario of
\cite{Fukugita:1986hr}.

Taking the minimum of the Higgs potential gives  
\begin{eqnarray}
\frac{dV}{d\phi} = 0 = -4a^2M^2 \phi + 4a^2 \phi^3
+2(n+1) \lambda^2m_{pl}^4 \frac{\phi^{2n+1}}{m_{pl}^{2n+2}}.
\label{dv}
\end{eqnarray}
Defining $a \equiv M/m_{pl}$, we are interested in the regime
$a,\lambda \ll 1$, for which the solution to
Eq. (\ref{dv}) is
\begin{equation}
\langle \phi^2_{\rm min} \rangle 
\approx M^2[1 - \frac{(n+1)}{2} \lambda^2 a^{2n-4}].
\end{equation}
At this minima
\begin{equation}
V_{\rm min} \equiv V(\langle \phi^2_{\rm min} \rangle) \approx
\lambda^2 a^{2(n+1)} m_{Pl}^4 ,
\end{equation}
\begin{equation}
m_{\phi}^2 \equiv V''(\phi_{\rm min}) \approx
8 a^2 M^2 , 
\end{equation}
and
\begin{equation}
V_0 \equiv V(\phi=0) = a^2 M^4 .
\end{equation}
Choosing the scale $M \sim 10^{10-11} {\rm GeV}$ leads to
\begin{eqnarray}
V_0^{1/4} & \approx & 10^{5-8} {\rm GeV} \nonumber \\
m_{\phi} & \approx & 10^{2-4} {\rm GeV} ,
\label{scales}
\end{eqnarray} 
which are the desired properties for the 
thermal inflation zero temperature potential.
Moreover, $\lambda$ remains a free parameter along with a
choice for the index $n$ of the higher dimensional operator.
This implies the value of $V_{\rm min}$ remains at our
discretion, and it can be chosen to
give the desired  scale of the presentday cosmological constant.  
In particular,
for $ V_{\rm min} = \rho_{\Lambda} \approx 10^{-10} eV^4$
it implies the condition $\lambda \approx 10^{-53 + 8n}$.
So, for example, for $n=2$ it requires $\lambda \approx 10^{-37}$
whereas for $n=6$, $\lambda \approx 10^{-5}$.
It is interesting that both the moduli and cosmological constant problems
can be solved by this model, but parametrically
neither of these two examples are particularly desirable. $\lambda$ must
be highly fine tuned in the $n=2$ case, and although $\lambda$ is a typical
coupling for the lepton sector of the standard model when $n=6$, the relevant
term in the O'Raifeartaigh model Higgs potential is of order $|\phi|^{14}$.
Another undesirable feature of the model in its present form is, since it
only respects global SUSY, after symmetry breaking for $\phi$,
since $V_{\rm min} \approx 0$, SUSY remains only 
very weakly broken \cite{witten} and so uninteresting
for particle physics.
Later we will propose a scenario where incorporating this model
within a local supersymmetric theory can overcome
all these problems, yet preserve those features
attractive for solving cosmological problems. 

At low-temperature the O'Raifeartaigh potential has the shape
and scales necessary for thermal inflation and at
$T=0$ its minima can be chosen to give the
scale of the presentday cosmological constant. 
{}For thermal inflation, it still must
be confirmed that at high temperature, $T > a^{1/2} M$, 
thermal corrections to the effective potential stabilize
$\phi$ at zero.  Lowest order finite temperature
corrections to SUSY models shift the mass as shown in 
\cite{de97}, and this argument can be modified
to the generalized O'Raifeartaigh models.  
Since $\lambda$ is tiny, the dominant
high temperature corrections will come in Eq. (\ref{genorpot}) from
the terms $a^2 \phi^4$ and $4a^2 \phi^2 \chi^2$,
which lead to high-T terms $\sim a^2T^2 \phi^2$ and
$\sim a^2 T^2 \chi^2$.  Thus, for $T> a^{1/2}M$ the minimum
of the effective potential will be as desired at
$<\phi>_T = < \chi>_T = <\eta>_T = 0$.

It is interesting to note that independent of the thermal
inflation problem, for an appropriate choice of scales,
the potential Eq. (\ref{genorpot}) can be implemented just
to address the cosmological constant problem.  In particular,
the minimum scale necessary to obtain adequate vacuum energy
is $M \stackrel{>}{\sim} \Lambda_{QCD}$.
An interesting case is when $M \sim 10^3 {\rm GeV}$,
the electroweak scale, where for the simplest 
nonrenormalizable potential
$n=2$,
\begin{equation}
V(\phi_{min} = M) = \lambda^2 10^{16} {\rm eV}^4 ,
\end{equation}
which is at the scale of $\rho_{\Lambda}$ for $\lambda \sim 10^{-13}$.
Moreover independent of $\lambda$, at the minimum
$m_{\phi} \equiv \sqrt{V''(\phi_{min})} \approx 10^{-(3-5)} {\rm eV}$.
This is an interesting scale as it is in the neighborhood
of the neutrino mass mixing parameters.

After thermal inflation, once the flaton $\phi$ is near its minimum,
it will oscillate and thereby enter a reheating phase similar
to that after supercooled inflation.  The particle production
history that develops has the same range of possibilities and
outcomes as studied in other thermal inflation works 
\cite{Lyth:1995hj,Lyth:1995ka,Lazarides:ja,Barreiro:1996dx,Asaka:1999xd}.
{}For example, if $\phi$ is a gauge singlet as in Eq. (\ref{genor}),
then reheating will create $\phi$-bosons.  Also,
$\phi$ can couple to charged scalars which can mediate decay into
gauge particles.  On the other hand, it is possible to
easily generalize our 
O'Raifeartaigh models by letting $\phi$ be in
the adjoint representation of some gauge group G, while
keeping $\chi$ and $\eta$ as singlets of G.  Thus, making
the replacement
$\phi^{p} \rightarrow {\rm Tr} (A^{p})$,
a nonvanishing VEV for $A$ can break G
to a set of degenerate minima, although gravity
will lift the degeneracy (see below).
{}For example for G = SU(N), $\langle A \rangle$ can 
be diagonalized by a SU(N) transformation
so G may break to subgroups of the form
\begin{equation}
H= \prod_i SU(N_i) \times U^p(1) ,
\end{equation}
where $\sum_i(N_i-1) + p = N-1$, i.e., H has the
same rank as G.  This form of O'Raifeartaigh models has more
possibilities of dissipating the vacuum energy.

The O'Raifeartaigh type models we have been discussing up
to now have global SUSY.  
In this case, the symmetry breaking considered above 
does not lead to SUSY breaking at scales of 
interest to particle physics, since the vacuum energy
at the minimum is essentially zero.
The full theory is expected 
to start off locally  supersymmetric, thus be a supergravity
theory.  It is well known that
global SUSY models can have many degenerate minima as long as
SUSY is unbroken.  SUGRA lifts this degeneracy \cite{Weinberg:id}
and only one minimum can have zero energy, with the others having
negative energy.  These results discussed in \cite{Weinberg:id}
were explicitly stated to exclude models of the 
O'Raifeartaigh \cite{O'Raifeartaigh:pr} and Fayet-Iliopoulos \cite{fi} type.
If O'Raifeartaigh potentials are included, they break SUSY and make positive 
contributions to vacuum energy.  It is thus quite possible
that one of the negative vacuum energy SUGRA minima receives
an additional positive vacuum energy contribution 
from the O'Raifeartaigh sector.  Thus, while 
both the (+) and ($-$) contributions are large, the 
{\it residual vacuum energy} can be small and positive.  This
could be the true vacuum energy
of the universe, and so explain the observed
cosmological constant.  {}For example consider the parameters
necessary for near balancing vacuum contributions
at a scale relevant to particle physics SUSY symmetry breaking,
$\sim 10^{3} {\rm GeV}$.  {}For the scale considered in
Eq. (\ref{scales}), $M \sim 10^{10-11} {\rm GeV}$, for $n=2$ 
to obtain $V_{\rm min}^{1/4} \stackrel{<}{\sim} 10^3 {\rm GeV}$,
it requires $\lambda \stackrel{<}{\sim} 10^{-5}$, which
is a realistic value. It remains a model building 
challenge to realize this effect through a natural mechanism.

It appears the inclusion of O'Raifeartaigh superpotentials in the full
SUGRA has interest for both particle physics and cosmology.
While breaking SUSY adequately to generate potentially
interesting phenomenological particle spectra, 
the O'Raifeartaigh potential
can also shift the vacuum to a small positive value, generating
the cosmological constant and from our above treatment,
the same model can solve the moduli problem by permitting
realization of thermal inflation.  This scenario is
promising and it seems worth further developing toward
a  realistic model. An initial step is to understand the
origin of such models from fundamental theories.
It is known that various compactifications of string theory
have a number of light scalar singlets in their spectrum.
{}For instance many models obtained from type IIB strings
via orbifolding AdS$_5 \times$ S$^5$ lead to such scalars.
The form of the superpotential is certainly model 
dependent. {}For example, it will depend on the initial string theory,
or more generally the initial region of parameter space in M-theory,
and details of the compactification.  However, the occurrence of scalars
are generic, so O'Raifeartaigh potentials can naturally arise in SUGRA
and thus lead to our cosmology.
 
To summarize, in this Letter we have shown that generalized O'Raifeartaigh
models can have powerful implications both for
cosmology and uniting cosmology with particle physics.
Within the compact structure of these models, we have 
shown that they can solve the moduli problem and
potentially lead to a solution of the 
cosmological constant problem.  In an attempt to unify
the symmetry breaking necessary to solve these cosmological
problems with that necessary to break SUSY in  particle physics
models, a new interpretation of the presentday accelerating
universe emerges providing dark matter and a ``balanced'' 
residual vacuum energy.  This is an
intriguing coincidence of solutions, given that O'Raifeartaigh 
type models may arise generically from fundamental theories.  

While we believe our scenario is provocative, more work needs to be done if
it is to be developed into a completely satisfying model. 
Some way to avoid fine tuning of positive and negative
contributions to the vacuum energy, or at least
the renormalization of the fine tuning parameters (This is already provided for in the superpotential
above the SUSY scale.) would be a important step.
Another avenue to follow would be to develop a similar scenario for 
Fayet-Iliopoulos \cite{fi} D-term SUSY breaking.

We thank Thomas Binoth and Michael Kr{\"a}mer for a useful
discussion.
A.B. is funded by the United Kingdom
Particle Physics and Astronomy Research Council (PPARC).
T.K. was partially supported by PPARC and U.S. DOE 
grant number DE-FG05-85ER40226 and
thanks the University of Edinburgh for its hospitality and 
partial support.

\end{document}